\documentclass[letterpaper]{amsart}
\usepackage{amssymb}

\newtheorem{theorem}{Theorem}

\theoremstyle{definition}
\newtheorem{definition}[theorem]{Definition}

\title[Sampleable hard instances]{A Finitely presented group whose word problem has sampleable hard instances}

\author{Robert H. Gilman}

\thanks{Partially supported by NSF Grant 1318716 }

\date{\today}

\begin{document}

\begin{abstract} Hard instances of natural computational problems are often elusive. In this note we present an example of a natural decision problem, the word problem for a certain finitely presented group, whose hard instances are easy to find. More precisely the problem has a complexity core sampleable in linear time.
\end{abstract}

\maketitle

\section{Introduction}
In 1975 Nancy Lynch~\cite{L} proved that for every computable decision problem not decidable in polynomial time there exists an infinite computable set of instances, 
$X$, such that the problem cannot be decided in polynomial time on any infinite subset of $X$. Such an $X$ is called a complexity core for the decision problem. 

Lynch's result attracted the attention of several other authors, who considered decision problems in the form of membership problems for subsets $S\subseteq \{0,1\}^*$. Cores of non-sparse density with membership decidable in subexponential time are investigated in~\cite{DB, OS}. If a core exists, then so does a proper core~\cite{ESY}, i.e., $X\subseteq S$. Proper cores are necessarily $\bf P$-immune, and $\{0,1\}^*$ is itself a core if and only if $S$ is $\bf P$-bi-immune~\cite{BS}. Generalizations to complexity classes beyond $\bf P$ are given in~\cite{BD1, BD2, ESY}. These results are reviewed in~\cite[Chapter 6]{BDG}.

Lynch's construction of cores involves enumeration of all Turing machines, and in general the membership problem for cores is superpolynomial. In~\cite{OS} the authors observe that as all known cores are more or less artificially constructed (when the core is $\{0,1\}^*$, it is the $\bf P$-bi-immune set $S$ which is artificially constructed), it would be extremely interesting to find natural examples of cores. Theorem~\ref{thm} exhibits such a core, albeit with respect to a slight variation of Lynch's original definition of cores. 

\subsection*{Notation} For any finite set $\Sigma$, $\Sigma^*$ is the set of all words over $\Sigma$, and $\hat\Sigma$ is the union of $\Sigma$ with a disjoint set of formal inverses.

\begin{theorem}\label{thm}
	There exists a finitely presented group $G =\langle \Sigma \mid R\rangle$ and a nonempty subset $\Delta\subset \Sigma$ such that if $D$ is the domain of convergence for any (correct) partial algorithm deciding the word problem, then
	\begin{equation}\label{eq}
	\lim_{n\to\infty}\frac{|D\cap \hat\Delta^n|}{|\hat\Delta^n|} =0
	\end{equation} 
	where $\hat\Delta^n$ denotes the set of all words of length at most $n$ in $\hat\Delta^*$.
\end{theorem}

The word problem for $G$ (with respect to the given presentation) is to decide whether an arbitrary word over $\hat\Sigma$ represents the identity in $G$. Theorem~\ref{thm} says that every partial algorithm for the word problem fails on virtually all words from $\hat\Delta^*$. Thus $\hat\Delta^*$ is a readily available set of provably hard instances. Clearly membership in $\hat\Delta^*$ is decidable in linear time, and $\hat\Delta^*$ can be sampled in linear time. In addition $\hat\Delta^*$ is a complexity core in the sense of Lynch except that the set inputs from $\hat\Delta^*$ on which a partial algorithm succeeds is not finite, as in Lynch's definition of a core,  but rather of asymptotic density zero in the sense of Equation~\ref{eq}. 

Theorem~\ref{thm} is an immediate consequence of Theorem~\ref{alg}, a recent result from combinatorial group theory. 

\section{Background and Proof}

Recall that in a finite presentation $\langle \Sigma \mid R\rangle$, $\Sigma$ is a finite set of generators and $R$ is a finite set of relators, i.e., of words over $\hat\Sigma$. Also an arbitrary word over $\hat\Sigma$ represents the identity of $G$ if and only if it can be reduced to the empty word by inserting and deleting words from $R$ and their inverses, along with the trivial words $aa^{-1}, a^{-1}a$ for $a\in\Sigma$. 

\begin{definition}[\cite{MO}] A finitely generated group $H$ is algorithmically finite if every infinite computably enumerable subset of words in the generators and their inverses contains two words which represent the same element of $H$.
\end{definition}

All finite groups are algorithmically finite. The interesting fact is that infinite algorithmically finite groups exist.

\begin{theorem} [\cite{MO} Theorems 1.1 and 1.3]\label{alg}
Infinite recursively presented algorithmically finite groups exist. Any partial algorithm for the word problem of such a group converges only on a set of asymptotic density zero. 
\end{theorem}

\subsection*{Proof of Theorem~\ref{thm}} Let $H$ be an infinite finitely generated recursively presented algorithmically finite group. By the well known Higman embedding Theorem $H$ is a subgroup of a finitely presented group $G$. Without loss of generality the generators $\Sigma$ of $G$ can be augmented to include generators, $\Delta$, of $H$. By Theorem~\ref{alg} any partial algorithm for the word problem of $G$ fails everywhere on $\Delta^*$ except on a subset of asymptotic density $0$ in $\Delta^*$. 

\section{Conclusion}

Infinite algorithmically finite groups are a new kind of group with unsolvable word problem. The proof of Theorem~\ref{alg} does not employ Turing machines; instead, Golod--Shafarevich presentations and analogs of simple sets from computability theory are used. 

The construction of $H$ is not natural in our sense, as it involves enumeration of all recursively enumerable subsets of $\hat\Sigma^*$. However $G$ itself is specified by a straightforward finite presentation. Since the proof of the Higman Embedding Theorem is constructive~\cite{R}, as is the construction of the recursive presentation for $H$, one could in principle compute this finite presentation.
 
The convergence of the limit in Theorem~\ref{thm} can be made to occur exponentially fast. See~\cite{MO} Corollary 1.4.

Sampleability of hard instances is of interest in crytography. The word problem of the group $G$ from Theorem~\ref{thm} is unsolvable and thus not useful as a cryptoprimitive. It seems unlikely that our approach could produce a useful cryptoprimitive, but examples of sampleable cores at lower complexity levels might provide some useful insights.

\end{document}